\documentclass[9pt, conference]{IEEEtran}
\IEEEoverridecommandlockouts
\usepackage{cite}
\usepackage{amsmath,amssymb,amsfonts}
\usepackage{algorithmic}
\usepackage{graphicx}
\usepackage{textcomp}
\usepackage{xcolor}
\usepackage{blindtext}
\usepackage{hyperref}
\usepackage{pgfplots}
\usepackage{xcolor}
\usepackage{booktabs} 
\usepackage{bbm}
\usepackage{bm}
\usepackage{multirow}
\usepackage{siunitx}
\def\BibTeX{{\rm B\kern-.05em{\sc i\kern-.025em b}\kern-.08em
    T\kern-.1667em\lower.7ex\hbox{E}\kern-.125emX}}
\begin{document}

\title{EMO-Debias: Benchmarking Gender Debiasing Techniques in Multi-Label Speech Emotion Recognition}


\author{
\IEEEauthorblockN{Yi-Cheng, Lin}
\IEEEauthorblockA{
    \textit{National Taiwan University}\\
    Taipei, Taiwan \\
    f12942075@ntu.edu.tw
}
\and
\IEEEauthorblockN{Huang-Cheng Chou}
\IEEEauthorblockA{
    \textit{Independent Researcher} \\
    Taipei, Taiwan \\
    huangchengchou@gmail.com
}
\and
\IEEEauthorblockN{Yu-Hsuan Li Liang}
\IEEEauthorblockA{    
    \textit{National Taiwan University}\\
    Taipei, Taiwan \\
    b10902112@csie.ntu.edu.tw}
\and
\IEEEauthorblockN{Hung-yi Lee}
\IEEEauthorblockA{
    \textit{National Taiwan University}\\
    Taipei, Taiwan \\
    hungyilee@ntu.edu.tw}
}

\maketitle

\begin{abstract}
Speech emotion recognition (SER) systems often exhibit gender bias.
However, the effectiveness and robustness of existing debiasing methods in such multi-label scenarios remain underexplored. 
To address this gap, we present \textbf{EMO-Debias}—a large-scale comparison of 13 debiasing methods applied to multi-label SER. 
Our study encompasses techniques from pre-processing, regularization, adversarial learning, biased learners, and distributionally robust optimization. 
Experiments conducted on acted and naturalistic emotion datasets, using WavLM and XLSR representations, evaluate each method under conditions of gender imbalance. 
Our analysis quantifies the trade-offs between fairness and accuracy, identifying which approaches consistently reduce gender performance gaps without compromising overall model performance. 
The findings provide actionable insights for selecting effective debiasing strategies and highlight the impact of dataset distributions.
\end{abstract}


\section{Introduction}
\emph{Speech Emotion Recognition} (SER) plays a vital role in human-centric AI applications, particularly in mental health monitoring, where early detection of emotional states can facilitate timely intervention \cite{WHO_2013, Kerkeni_2019, Elsayed_2022, Adeleye_2024}. However, studies \cite{Lin_2024, Gorrostieta_2019, lin24b_interspeech} have consistently found gender bias in deep learning-based SER systems, often favoring female speakers over males. Such performance bias is especially problematic in mental health applications, where misclassification of male emotional states could delay critical interventions.
Several strategies have been proposed to mitigate bias in speech processing systems \cite{lin2025mitigatingsubgroupdisparitiesmultilabel, zhang2023exploringdataaugmentationbias, 10430478}. Gorrostieta et al. \cite{Gorrostieta_2019} was the first to assess gender bias on dimensional SER (e.g., valence or arousal) models, and a recent study \cite{Feng_2022} on single-label categorical SER has also tackled this issue.

Despite increasing recognition of gender bias in SER \cite{Chien_2023, Chien_2024, Upadhyay_2025, Lin_2024}, research has not yet fully examined debiasing techniques. 
Most existing debiasing techniques, which have primarily been developed for single-label classification in computer vision and natural language processing, have not been systematically evaluated in multi-label scenarios. 
With the emerging shift from single-label to multi-label SER \cite{Chou_2022,open_emotion}, it is essential to investigate whether these methods can effectively mitigate bias while handling the added complexity of multi-label emotion predictions.

To address existing gaps in the field, we present \textbf{EMO-Debias}, the first large-scale evaluation of debiasing methods for multi-label SER. 
We adapt 12 established techniques and propose one novel method, including adversarial training, biased learner, regularization, data pre-processing, and distribution robust optimization, to mitigate gender bias while maintaining robust multi-label SER performance.
We validate these methods using two public emotion datasets, the MSP-Podcast \cite{Lotfian_2019} and CREMA-D \cite{Cao_2014}. 
Also, we simulate controlled gender imbalances in the training data, ranging from a balanced 1:1 ratio to a highly skewed 1:40 ratio. 
This allows us to systematically examine the impact of data distribution on model performance and fairness.
To sum up, our key contributions are as follows:
\begin{enumerate} 
\item \textbf{Comprehensive Benchmark}: 
We introduce \textbf{EMO-Debias}, the first large-scale benchmark of 13 debiasing methods for multi-label SER, providing guidance for selecting effective and robust debiasing strategies.
\item \textbf{Bias Impact Analysis}: 
We empirically demonstrate how gender distribution imbalances impact both the accuracy and bias of SER models.
\item \textbf{Novel Adaptation}: 
We adapt twelve single-label debiasing techniques for the multi-label framework that effectively mitigates gender bias in complex multi-label scenarios.
\end{enumerate}

By addressing gender bias in multi-label SER, our research advances the broader goal of developing fair, transparent, and inclusive AI systems. Our findings offer practical insights for mitigating bias in SER models. To ensure reproducibility, all code and datasets will be made publicly available upon paper acceptance.

\section{Background and Related Works}
\subsection{Multi-label SER}
Emotion perception is inherently complex, and speakers often convey multiple emotions at once. 
However, most previous studies have treated categorical SER as a single-label task, overlooking the ambiguity of emotions \cite{Wu_2024_v2, Chou_2024_v3}. 
Since an utterance might be both angry and sarcastic, multi-label classification is more suitable for SER.
Following recent SER research \cite{Chou_2024_v3, open_emotion, park2024multi} and psychological insights \cite{Cowen_2017, Cowen_2021}, we define SER as a multi-label task to better reflect the complexity of emotion perception.

\subsection{Gender Debiasing on SER Task}

While studies like \cite{Lin_2024, Gorrostieta_2019, lin24b_interspeech} have identified gender bias in SER models, only a few have proposed debiasing methods. 
The pioneering work by \cite{Gorrostieta_2019} introduced a two-stage debiasing training method using adversarial learning to enhance model fairness, resulting in more balanced accuracy across genders. 
Chou et al. \cite{Chou_2024} applied unsupervised clustering of utterance embeddings from a pre-trained speaker verification model to estimate speaker identities and apply fairness constraints without prior knowledge of speaker IDs. 
Their work, however, targets emotion regression rather than categorical recognition. In contrast, the present study focuses on multi‐label categorical SER.


Recent efforts \cite{Chien_2023, Chien_2024, Upadhyay_2025} aim to improve fairness in single‐label SER by accounting for annotator and speaker gender. These approaches often limit the emotion set (for example, using only four primary emotions from MSP‐PODCAST \cite{Lotfian_2019} instead of all eight). Such practices can skew evaluation and restrict the claimed benefits of debiasing. To provide a more realistic assessment, this work uses the complete original emotion pools, framing SER as a multi‐label task.

\section{Dataset}
This study utilizes two publicly available emotion databases, the MSP-PODCAST and CREMA-D, which represent naturalistic (real-world) and acted emotions in English, respectively, to ensure comprehensive experimental validation.

\textbf{The MSP-PODCAST} is the largest public naturalistic emotion database segmented from real-world podcast recordings \cite{Lotfian_2019}. We use its v1.12 release, comprising 324.38 hours of speech from 3,513 speakers. The dataset is partitioned into five subsets: Train, Development, and Test1–Test3. We merge Test1 and Test2 into a single Test set and exclude Test3 due to unavailable labels \cite{emo_superb, open_emotion, Chou_2024_v4}. We then filter the test set samples to keep those that contain a gender label and a dominant emotion class (i.e., one emotion class has a label distribution exceeding 0.5), resulting in 33,873 test samples. This study focuses on eight primary emotions—angry, sad, happy, surprise, fear, disgust, contempt, and neutral—framing the task as an 8-class emotion classification.



\textbf{The CREMA-D} \cite{Cao_2014} comprises 7,442 utterances from 91 actors delivering 12 sentences in six emotions: anger, disgust, fear, happy, neutral, and sad. 
The dataset is annotated by at least seven raters in audio-only, visual-only, and audio-visual settings. This study uses audio-only ratings, as suggested in \cite{Chou_2025}. As there are no official partitions, we follow the EMO-SUPERB \cite{open_emotion} partitioning to perform cross-fold validation for consistency and reproducibility.

\begin{table}[t]
\centering
\caption{Bias-amplified data distribution (1:20 ratio) in MSP-PODCAST and CREMA-D dataset (Fold 1). The numbers are utterance counts. F: Female; M: Male; Dev: Development set.}
\vspace{-2mm}
\label{tab:data_distribution}
\begin{tabular}{lcrrrrrr}
\toprule
\multirow{2}{*}{Emotion} & \multirow{2}{*}{Gender} 
  & \multicolumn{3}{c}{MSP-PODCAST} & \multicolumn{3}{c}{CREMA-D} \\
\cmidrule(lr){3-5} \cmidrule(lr){6-8}
& &Train & Dev & Test &Train & Dev & Test \\
\midrule
Angry    & F &   167  &  142 & 2709 &  12  &  4  & 102 \\
         & M &  3357  & 2845 & 2452 & 240  & 93  &  58 \\ \midrule
Disgust  & F &    11  &    5 &  170 &   4  &  1  &  41 \\
         & M &   235  &  110 &  149 &  86  & 34  &  18 \\ \midrule 
Neutral  & F &   688  &  149 & 6750 &  54  & 19  & 263 \\
         & M & 13771  & 2980 & 7022 & 1080 & 384 & 250 \\ \midrule
Fear     & F &   218  &  158 &  126 & 127  & 27  &  55 \\
         & M &    10  &    7 &  150 &   6  &  1  &  20 \\ \midrule
Happy    & F &   357  &  154 & 5901 & 102  & 24  &  35 \\
         & M &  7142  & 3083 & 4616 &  5  &  1  &  15 \\ \midrule
Sad      & F &  2051  &  963 & 1226 & 58  & 41  &  35 \\
         & M &   102  &   48 & 1148 &  2  &  2  &   7 \\ \midrule
Surprise & F &   576  &  230 &  497 &\multicolumn{3}{c}{--} \\
         & M &    28  &   11 &  403 &\multicolumn{3}{c}{--} \\ \midrule
Contempt & F &   296  &  490 &  369 &\multicolumn{3}{c}{--} \\
         & M &    14  &   24 &  185 &\multicolumn{3}{c}{--} \\
\bottomrule
\end{tabular}
\label{tab:1_20_data_distribution}
\end{table}


\section{Baseline SER Framework and Evaluation}

\subsection{\textbf{Task Definition}}
\label{subsec: task_definition}
For each utterance \(x\), we have a distributional emotion label \(y\) and a speaker gender \(g\). Our model is trained using a frozen self-supervised learning (SSL) backbone and a weighted-sum strategy. We encode the raw speech \(x\) with the SSL encoder \(E_S\) to obtain a hidden representation \( h_{S} = E_S(x) \). Subsequently, we feed \(h_S\) into a linear classifier \(C\) to predict the \textbf{emotion distribution} \(\hat{y} \;=\; C\bigl(h_S\bigr)\).

\subsection{\textbf{Data Pre-processing}}
\label{subsec: data_pre-process}
This study investigates debiasing methods in multi-label SER.
Following \cite{Riera_2019, Chou_2024_v2}, distributional labels are calculated based on the frequency of emotional ratings and used for training and evaluating SER systems. In line with \cite{Chien_2023, Chien_2024, Upadhyay_2025}, we then select only those samples in which one emotion class receives over half of the annotators' votes. 

Real‐world speech datasets often exhibit nontrivial gender imbalances. For example, early Common Voice English subset reports a male:female speaker ratio exceeding 9:1 \cite{common_voice}. In TIMIT \cite{timit}, there are 326 male versus 136 female speakers (roughly 2.4:1). To evaluate debiasing under both typical and stress‐test conditions, training and development sets are adjusted to ratios of 1:1, 1:5, 1:10, 1:20, and 1:40, while the test set remains unchanged.

To introduce bias, we adjust the ratios between male and female utterances, considering original emotion distributions. For instance, in the MSP-PODCAST corpus, anger, disgust, happiness, and neutral are more common in male speakers, while contempt, fear, sadness, and surprise are prevalent among female speakers. We amplify this bias by modifying the male-to-female ratios in the training and development sets. Table~\ref{tab:1_20_data_distribution} presents the detailed data distribution at a 1:20 ratio, using Fold 1 of the CREMA-D and the MSP-PODCAST datasets as examples.


\subsection{\textbf{SER Classifier}}
All experiments in the work follow the EMO-SUPERB framework \cite{open_emotion}, based on the S3PRL recipe \cite{Yang_2024}, using WavLM\footnote{\tiny https://huggingface.co/s3prl/converted\_ckpts/resolve/main/wavlm\_base\_plus.pt} and XLSR\footnote{\tiny https://dl.fbaipublicfiles.com/fairseq/wav2vec/xlsr2\_300m.pt}—the top SER leaderboard models—as fixed SSL backbones with 2 linear layer prediction head.  
The AdamW optimizer \cite{Loshchilov_2019} is applied with a learning rate of 1e-4 for WavLM and 5e-4 for XLSR, chosen after a learning‐rate search that yielded the best performance, and a batch size of 32.

\subsection{\textbf{Evaluation Metrics}}
\label{subsec:evaluation_metrics}

\subsubsection{Accuracy}
We use macro-F1 score (\textbf{F1}) and Hamming accuracy (\textbf{ACC})  as the SER performance metrics \cite{duret24_odyssey, 10887615}. \textbf{F1} accounts for class imbalances by computing the F1-score for each emotion category separately and averaging them, ensuring a fair assessment across all classes. In contrast, \textbf{ACC} measures the proportion of correctly classified labels across all predictions, providing an overall performance metric. 
We adopt the threshold, $1/C$, as used in \cite{open_emotion}, where $C$ represents the number of emotion classes, to binarize distributional labels for accuracy evaluation.

\subsubsection{Fairness}
We employ two primary criteria: Equalized Odds-based metrics \cite{Hardt_2016} and Demographic Parity (\textbf{DP$_{gap}$}). The Equalized Odds metrics are calculated as the root mean square (\textbf{RMS}) of the gaps in true-positive rate (\textbf{TPR$_{gap}$}) \cite{Han_2021} and false-positive rate (\textbf{FPR$_{gap}$}) \cite{jeong2022gets, tian2024softprompttuninglargelanguage} across all emotions, following \cite{Shi_2024}. These metrics ensure that predictions are independent of gender given the true outcome, thereby promoting similar TPR and FPR across demographic groups. The RMS of gaps in F1 score (\textbf{F1$_{gap}$}) further balances precision and recall to provide a comprehensive measure of bias.
\textbf{DP$_{gap}$} requires that a model’s predictions be independent of demographic groups, meaning that the probability of receiving a positive prediction should be equal across demographic groups, regardless of actual outcomes. Specifically, DP$_{gap}$ is calculated as the difference in the probability of a positive prediction between male and female groups:
\begin{equation} 
DP_{gap} = \sqrt{\sum_{\hat{y}\in\mathcal{Y}}\max_{g\in\mathcal{G}} DP(g, \hat{y})^2};
\end{equation}
\begin{equation} 
DP(g, \hat{y}) = \mathbb{E} \Big[P(\hat{y} = 1 \mid g) - P(\hat{y} = 1)\Big],
\end{equation} 
where \(\mathcal{G}\) represents the set of all predefined groups (genders) in the dataset. A high DP$_{gap}$ suggests that one gender receives significantly more positive classifications, indicating a potential bias in the model.


\section{Methodology}

This section presents our methodology for mitigating gender bias in multi-label SER. To ensure broad coverage of the major design paradigms in fairness research, we deliberately selected \emph{one or two representative algorithms from each of five canonical families}.
We adapt 12 existing debiasing methods, originally designed for single-label paradigms, to function effectively within our multi-label framework. 
Furthermore, we introduce a novel debiasing approach: Gap Regularization, crafted for the multi-label SER context. This balanced test bed (i) reflects approaches most frequently cited or state‐of‐the‐art in fairness and (ii) allows like‐for‐like comparison within each family. 
For each method, we will delineate the specific modifications, architectural adjustments, and algorithmic considerations undertaken. 
Table~\ref{tab:debias_methods} summarizes the objective functions of the adapted debiasing methods. 
We follow \cite{open_emotion, Chou_2024_v3} to use the class-balance cross-entropy loss (CE) \cite{Cui_2019} as the base loss function in the following sections.

In our adaptation to methods in Sec.~\ref{subsec: adverserial}, ~\ref{subsec: pre_processing}, and ~\ref{subsec: gdro}, we replace the original one-hot emotion targets with distributional labels (as described in Sec.~\ref{subsec: data_pre-process}), so that the emotion loss is computed against soft labels (distributions) rather than hard labels. Notice that those debiasing methods can be applied in the same way as in the single-label setting.

\begin{table}[t]
    \centering
    \fontsize{7.5}{9}\selectfont
    \setlength{\tabcolsep}{1pt}
    \caption{Summary of Debiasing Methods. \textbf{BS} refers to the debias methods that require \textbf{Bias Supervision}.}
    \vspace{-2mm}
    \begin{tabular}{@{\hspace{0cm}}l@{\hspace{0.04cm}}@{\hspace{0.04cm}}l@{\hspace{0.04cm}}@{\hspace{0cm}}c@{\hspace{0.01cm}}@{\hspace{0.01cm}}} 
        \toprule
        \textbf{Method} & \textbf{Objective Function} & \textbf{BS}  \\
        \midrule
        \textbf{ADV} \cite{single_adversary_1, single_adversary_2}  & $CE(y,\hat{y}) - \lambda_{adv} BCE(g,\hat{g})$ & \checkmark   \\
        \textbf{MADV} \cite{han-etal-2021-diverse}  & $CE(y,\hat{y}) - \frac{\lambda_{adv}}{k} \sum_k BCE(g,\hat{g}) + \lambda_{diff}\mathcal{L}_{diff}$ & \checkmark  \\
        \textbf{GR} [Ours]    & $CE(y,\hat{y})+\lambda_{GR}($TPR$_{gap}+$FPR$_{gap})$ & \checkmark \\
        \textbf{DS} \cite{kamiran2012data}   & $CE(y,\hat{y})$ & \checkmark  \\
        \textbf{RW} \cite{kamiran2012data}    & Sample weight adjusted by attribute frequency & \checkmark  \\
        \textbf{BLIND+d} \cite{orgad-belinkov-2023-blind} & $(1 - \sigma(f_B(x)))^{\gamma} CE(y,\hat{y}) + \lambda_B BCE(f_B(x), g)$& \checkmark   \\ 
        \textbf{GDRO} \cite{gdro}  & $\max_{g\in\mathcal{G}}(\mathbb{E}(CE(y, \hat{y})))$ & \checkmark   \\
        \textbf{GADRO} \cite{sagawadistributionally} & $\max_{g\in\mathcal{G}}(\mathbb{E}(CE(y,\hat{y}))+\frac{\lambda}{\sqrt{n_g}})$ & \checkmark  \\
        \midrule
        \textbf{LfF} \cite{Nam_2020}   & $W(x) CE(y,\hat{y}_D) + GCE(y,\hat{y}_B)$ &   \\
        \textbf{SiH} \cite{vandenhirtz2023signalisharder}   & $(1 - \hat{y}_B)^r CE(\hat{y}_D, y)+ GCE(y, \hat{y}_B)$ &   \\
        \textbf{DisEnt} \cite{lee2021learning} & $W(x) CE(y,\hat{y}_D) + GCE(y,\hat{y}_B) + \mathcal{L}_{swap}$ &    \\
        \textbf{BLIND-d} \cite{orgad-belinkov-2023-blind} & $(1 - \sigma(f_B(x)))^{\gamma} CE(y,\hat{y}) + \lambda_B BCE(f_B(x), ACC)$&  \\ 
        \textbf{LVR} \cite{LVR}   & $CE(y,\hat{y}) + \lambda_{LVR}\mathcal{L}_r+\mathcal{L}_c$ &   \\
        \bottomrule
    \end{tabular}
    \label{tab:debias_methods}
\end{table}

\subsection{\textbf{Adversarial Approaches}}
\label{subsec: adverserial}
We first obtain a hidden representation \(h_{S} = E_S(x)\) using the SSL encoder \(E_S\) (see Sec.~\ref{subsec: task_definition}). This \(h_{S}\) serves as the input for both the emotion classifier and the adversary networks. The emotion classifier \(C\) produces an emotion distribution \(\hat{y} = C(h_{S})\), while the adversaries operate on \(h_{S}\) to remove or obfuscate gender information.

\subsubsection{\textbf{Single adversary (ADV)}}
\textbf{ADV} \cite{single_adversary_1, single_adversary_2} introduces an adversarial encoder \(E_A\) and an adversarial classifier \(C_A\) for gender prediction. First, we compute \(h_{S} = E_S(x)\) using the SSL encoder. Then \(h_{A} = E_A(h_{S})\) is passed to \(C_A\) to predict speaker gender \(\hat{g} = C_A(h_{A})\). During training, \(E_A\) and \(C_A\) are optimized to minimize the gender prediction loss \(BCE(g, \hat{g})\), while \(E_S\) is updated to maximize this loss (i.e., to confuse \(C_A\)) and simultaneously minimize the emotion classification loss.

\subsubsection{\textbf{Multiple adversaries (MADV)}}
\textbf{MADV} \cite{Han_2021} extend the single-adversary framework by introducing \(k = 3\) adversarial encoders \(\{E_{A_i}\}_{i=1}^k\) and corresponding adversarial classifiers \(\{C_{A_i}\}_{i=1}^k\). During training, each \((E_{A_i}, C_{A_i})\) pair is optimized to minimize the binary cross-entropy loss \(BCE(g, \hat{g}_i)\), while \(E_S\) is simultaneously updated to maximize each gender loss (i.e., to confuse all \(C_{A_i}\)) and minimize the emotion classification loss. 
To ensure different discriminators capture diverse information, an additional difference loss $\mathcal{L}_{diff}$ is introduced, promoting orthogonality among the hidden representations of different sub-discriminators. $\mathcal{L}_{diff}$ is calculated by:
\begin{equation}
  \mathcal{L}_{diff} = \lambda_{diff} \sum_{i,j \in \{1,2,...,k\}} ||h_{A_i}^T h_{A_j}||^2 \mathbbm{1}\{i \neq j\},  
\end{equation}
where \( h_{A_i} \) is the hidden representation from the encoder \( E_{A_i} \), computed as \( h_{A_i} = E_{A_i}(h_{S}) \). \( ||\cdot||_F \) is the Frobenius norm, which measures the “size” of a matrix.  \( \lambda_{diff} \) is a hyperparameter that controls the strength of orthogonality regularization. The summation runs over all distinct pairs \((i, j)\) with \( i \neq j \), encouraging their representations to be as orthogonal (i.e., non-redundant) as possible. We set \(\lambda_{adv}=3.2\) and \(\lambda_{diff}=0.2\) in this work.

\subsection{\textbf{Pre-processing methods}}
\label{subsec: pre_processing}
Pre-processing methods change the sampling strategy or weight of samples during training. 

\subsubsection{\textbf{Reweighting (RW)}}
\textbf{RW} \cite{kamiran2012data} changes the weight of samples by the frequency of gender. Samples from underrepresented groups receive higher weights, ensuring that the model does not disproportionately favor the majority group.

\subsubsection{\textbf{Downsample (DS)}}
\textbf{DS} \cite{kamiran2012data} modifies the training dataset by resampling it to ensure that each emotion class contains an equal number of samples across all gender categories. This prevents the model from being biased toward attribute groups that are overrepresented in the original dataset.

\subsection{\textbf{Group distribution robust optimization}}
\label{subsec: gdro}
\subsubsection{\textbf{Group distribution robust optimization (GDRO)}}
\textbf{GDRO} \cite{gdro} trains model by minimizing the worst-group loss. Instead of optimizing for the average loss across all samples, GDRO focuses on the group with the highest loss to ensure that the model performs well even in the most challenging cases. The objective function for GDRO is defined in Table~\ref{tab:debias_methods}.

\subsubsection{\textbf{Group adjusted DRO (GADRO)}}

While GDRO effectively targets the worst-performing group, \textbf{GADRO} \cite{sagawadistributionally} highlights that directly applying GDRO on neural networks might fail on worst-case generalization because of vanishing worst-case training loss. Hence, a regularization term \(\frac{\lambda_{GD}}{\sqrt{n_g}}\) is proposed to prevent overfitting in smaller groups. Here \(\lambda_{GD}\) is a hyperparameter and \(n_g\) is the group size. We set \(\lambda_{GD}\) to 4 for the CREMA-D dataset and 20 for the MSP-PODCAST dataset.

\subsection{\textbf{Biased Learners}}
\label{subsec: biased_learner}
Biased learner approaches mitigate bias by training an auxiliary model to identify potentially biased samples without explicit bias labels. Bias labels refer to annotations of attributes such as gender, age, accent, or other characteristics that can introduce unwanted correlations. Since these labels are not provided (\textbf{no Bias Supervision (BS)}), the auxiliary model must learn to detect samples that rely on those attributes. Once identified, these samples are down-weighted so that the main model can focus on more challenging examples where bias does not dominate.

\subsubsection{\textbf{Learning from failure (LfF)}}
\textbf{LfF} \cite{Nam_2020} learns a biased model with the same architecture as the main model, using generalized cross-entropy (GCE) loss \cite{GCE}. The intuition is that the gradient of GCE loss up-weights the gradient of CE loss for samples with a higher predicted probability \( \hat{y}\). When a sample has a single label \(y_j\), the GCE loss and its gradient can be expressed as:
\begin{equation}
    GCE(y_j,\hat{y})=\frac{1-\hat{y}_j^q}{q} \Rightarrow \frac{\partial GCE(y, \hat{y})}{\partial \theta} = \hat{y}_j^q \frac{\partial CE(y, \hat{y})}{\partial \theta},
\end{equation}
where $\hat{y}_j$ is the predicted probability that the sample belongs to class $j$, \(q\) is a hyperparameter controlling the strength of reweighting. 
Following prior work \cite{GCE}, we set \(q\) to 0.7.
In our adaptation, instead of treating each class as strictly present or absent, we weight the GCE loss by the ground-truth probability \(y_j\) for each class. This yields:
\begin{equation}
GCE(y, \hat{y})=\sum_j{y_j\frac{1-\hat{y}_j^q}{q}}.
\end{equation}

Since \(y_j\) ranges between 0 and 1, the loss smoothly reflects how strongly each emotion is present. High-confidence predictions \(\hat{y}_j\) still receive larger gradient weights through the factor \(\hat{y}_j^q\), but now each term is proportionally scaled by the distributional label \(y_j\).

While training the biased model, a debiased model is also trained with the weight of the relative difficulty score:
\begin{equation}
W(x) = \frac{CE(y, \hat{y}_B)}{CE(y, \hat{y}_B)+CE(y, \hat{y}_D)},
\end{equation}
where \(\hat{y}_B\) and \(\hat{y}_D\) are the softmax logits of the biased and debiased model, respectively. Samples that align well with the biased model have a low $CE(y, \hat{y}_B)$, which makes $W(x)$ small.
This encourages the debiased model to focus more on difficult, bias-conflicting examples, improving its generalization across diverse inputs.

For robust LfF training, we compute the relative difficulty score using an exponential moving average of the cross-entropy losses 
\(\mathrm{CE}(y, \hat{y}_B)\) and \( \mathrm{CE}(y, \hat{y}_D)\),
instead of using the loss from each training epoch directly. We use a fixed exponential decay factor of \(\alpha = 0.7\), consistent with the original paper.

\subsubsection{\textbf{Signal is harder to learn than bias (SiH)}}
\textbf{SiH} \cite{vandenhirtz2023signalisharder} also trains a biased model by GCE loss. However, it trains the debiased model by detaching the logits from the biased model as in Table~\ref{tab:debias_methods},
where \(r\in(0, 1]\) is a hyperparameter set to control the strength of emphasis. We follow the original work \cite{vandenhirtz2023signalisharder}, set \(r=0.7\). With this loss, bias-conflicting samples are harder to learn than bias-aligned samples.
In our multi-label adaptation, we use the adapted multi-label GCE loss. For the debiased model, we apply a per-class reweighting factor \((1 - \hat{y}_{B,j})^{r}\) to each dimension of the multi-label CE loss: 
\begin{equation}
    (1 - \hat{y}_B)^{r} \,CE(y, \hat{y}_D)
\;=\; \sum_{j} \bigl(1 - \hat{y}_{B,j}\bigr)^{r} \, y_j \,\bigl[-\log(\hat{y}_{D,j})\bigr].
\end{equation}

By summing across all \(j\), the adapted debiased model focuses more on emotions where the biased model’s confidence is low (bias-conflicting), while still respecting the soft labels \(y_j\).

\subsubsection{\textbf{Disentangled Feature Augmentation (DisEnt)}}
\textbf{DisEnt} \cite{lee2021learning} trains two separate encoders to disentangle intrinsic attributes (which inherently define an emotion) from bias attributes (peripheral features correlated with the labels). The key innovation is a feature-swapping mechanism, where the intrinsic attributes from different samples are combined with biased attributes in a way that breaks their original correlation. The adapted multi-label GCE loss is also used for the biased model here.

\subsubsection{\textbf{BLIND(+/-d)}}
\textbf{BLIND} \cite{orgad-belinkov-2023-blind} offers two variants. In the demographics-aware version (denoted as \textbf{BLIND+d}), an auxiliary model, demographic detector (\(f_B\)), is trained alongside the main model to predict demographic attributes (e.g., gender); if successful, the corresponding sample is down-weighted. 

The demographics-free version \textbf{(BLIND-d)} instead trains a success detector (\(f_B\)) to predict whether the main model will classify a given sample correctly. If \(f_B\) predicts high confidence in a correct prediction, the sample is likely relying on biased shortcuts. We adapt the original success detector from single-label detection to multi-label detection by predicting the Hamming accuracy (\textbf{ACC}) of the main model. By predicting \textbf{ACC} instead of single-label correctness, the success detector learns to identify samples where the model relies on bias to “easily” get many labels correct simultaneously. This adaptation allows the model to handle complex multi-label dependencies.
We set $\gamma =0.7$ and $\lambda_{B}=1$ for the loss in Table~\ref{tab:debias_methods}.

\begin{table*}[t]
  \centering
  \sisetup{table-format=1.2}
  \caption{Effect of increasing gender‐imbalance on 8‐class SER performance using WavLM upstream. $\uparrow$  indicates that higher values correspond to better performance and lower values correspond to worse performance, while $\downarrow$ represents the opposite.}
  \vspace{-2mm}
  \label{tab:detailed_ratios_results_swapped}
  \begin{tabular}{l *{6}{S} *{6}{S}}
    \toprule
    \multirow{2}{*}{\textbf{Ratio}}
      & \multicolumn{6}{c}{\textbf{MSP-PODCAST}}
      & \multicolumn{6}{c}{\textbf{CREMA-D}} \\
    \cmidrule(lr){2-7} \cmidrule(lr){8-13}
      & {F1$\uparrow$} & {ACC$\uparrow$} & {TPR$_\mathrm{gap}$$\downarrow$} & {FPR$_\mathrm{gap}$$\downarrow$} & {F1$_\mathrm{gap}$$\downarrow$} & {DP$_\mathrm{gap}$$\downarrow$}
      & {F1$\uparrow$} & {ACC$\uparrow$} & {TPR$_\mathrm{gap}$$\downarrow$} & {FPR$_\mathrm{gap}$$\downarrow$} & {F1$_\mathrm{gap}$$\downarrow$} & {DP$_\mathrm{gap}$$\downarrow$} \\
    \midrule
    1{:}1   & 0.41 & 0.75 & 0.08 & 0.05 & 0.08 & 0.08  & 0.68 & 0.84 & 0.10 & 0.05 & 0.07 & 0.04 \\
    1{:}5   & 0.41 & 0.75 & 0.13 & 0.10 & 0.11 & 0.09  & 0.67 & 0.84 & 0.17 & 0.08 & 0.09 & 0.06 \\
    1{:}10  & 0.41 & 0.73 & 0.17 & 0.13 & 0.12 & 0.11  & 0.66 & 0.83 & 0.24 & 0.11 & 0.11 & 0.09 \\
    1{:}20  & 0.43 & 0.72 & 0.19 & 0.16 & 0.13 & 0.12  & 0.65 & 0.82 & 0.28 & 0.13 & 0.12 & 0.10 \\
    1{:}40  & 0.42 & 0.72 & 0.21 & 0.17 & 0.13 & 0.14  & 0.64 & 0.82 & 0.31 & 0.16 & 0.12 & 0.12 \\
    \bottomrule
  \end{tabular}
  \label{tab:detailed_ratios_results}
\end{table*}

\begin{table*}[t]
  \centering
  \sisetup{table-format=1.2}
  \caption{Overall performance of baseline and de‐bias methods on the MSP-PODCAST and CREMA-D using \textbf{WavLM upstream}. In each column, the best performance is shown in \textbf{bold} and the second‐best is \underline{underlined}.}
  \vspace{-2mm}
  \label{tab:debias_wavlm}
  \begin{tabular}{l c *{6}{S} *{6}{S}}
    \toprule
    \multirow{2}{*}{\textbf{Method}} 
      & \multirow{2}{*}{\textbf{BS}} 
      & \multicolumn{6}{c}{\textbf{MSP-PODCAST}} 
      & \multicolumn{6}{c}{\textbf{CREMA-D}} \\
    \cmidrule(lr){3-8} \cmidrule(lr){9-14}
      &  & {F1$\uparrow$} & {ACC$\uparrow$} & {TPR$_\mathrm{gap}$$\downarrow$} & {FPR$_\mathrm{gap}$$\downarrow$} & {F1$_\mathrm{gap}$$\downarrow$} & {DP$_\mathrm{gap}$$\downarrow$}
         & {F1$\uparrow$} & {ACC$\uparrow$} & {TPR$_\mathrm{gap}$$\downarrow$} & {FPR$_\mathrm{gap}$$\downarrow$} & {F1$_\mathrm{gap}$$\downarrow$} & {DP$_\mathrm{gap}$$\downarrow$} \\
    \midrule
    ERM       & --          & 0.43 & 0.72 & 0.19 & 0.16 & 0.13 & 0.12  & 0.65 & 0.82 & 0.28 & 0.13 & 0.12 & 0.10 \\ \midrule
    ADV       & \checkmark  & 0.39 & \textbf{0.75} & 0.52 & 0.29 & 0.48 & \underline{0.03}
                          & 0.64 & 0.82 & 0.64 & 0.19 & 0.66 & \underline{0.04} \\
    MADV      & \checkmark  & 0.41 & 0.75 & 0.52 & 0.28 & 0.49 & 0.04
                          & 0.64 & 0.82 & 0.65 & 0.19 & 0.66 & 0.05 \\
    GR        & \checkmark  & \textbf{0.42} & 0.73 & 0.17 & 0.14 & 0.13 & 0.11
                          & \underline{0.65} & \underline{0.82} & 0.27 & 0.13 & 0.12 & 0.10 \\
    DS        & \checkmark  & 0.35 & 0.75 & \underline{0.09} & \underline{0.06} & \textbf{0.08} & \underline{0.09}
                          & 0.60 & 0.80 & \textbf{0.09} & \textbf{0.07} & \textbf{0.06} & \textbf{0.03} \\
    RW        & \checkmark  & 0.37 & \underline{0.75} & \textbf{0.08} & \textbf{0.06} & \underline{0.09} & \textbf{0.08}
                          & \textbf{0.65} & \textbf{0.82} & \underline{0.17} & \underline{0.08} & \underline{0.10} & 0.06 \\
    BLIND+d   & \checkmark  & \underline{0.41} & 0.72 & 0.17 & 0.15 & 0.11 & 0.12
                          & 0.57 & 0.79 & 0.32 & 0.17 & 0.18 & 0.13 \\
    GDRO      & \checkmark  & 0.42 & 0.68 & 0.19 & 0.16 & 0.10 & 0.13
                          & 0.63 & 0.80 & 0.20 & 0.12 & 0.08 & 0.09 \\
    GADRO     & \checkmark  & 0.42 & 0.68 & 0.16 & 0.12 & 0.10 & 0.11
                          & 0.62 & 0.79 & 0.19 & 0.13 & 0.08 & 0.08 \\

    \midrule
    BLIND-d   & --          & 0.42 & 0.72 & 0.20 & 0.16 & 0.13 & 0.13
                          & 0.64 & 0.82 & \underline{0.24} & \underline{0.11} & 0.13 & \underline{0.09} \\
    LfF       & --          & 0.42 & \underline{0.73} & 0.18 & 0.15 & \textbf{0.12} & 0.12
                          & \textbf{0.66} & \underline{0.83} & 0.27 & 0.13 & \underline{0.12} & 0.10 \\
    LVR       & --          & \underline{0.43} & 0.72 & 0.18 & \textbf{0.13} & 0.13 & \textbf{0.11}
                          & 0.64 & 0.81 & \textbf{0.21} & \textbf{0.09} & \textbf{0.10} & \textbf{0.08} \\
    SiH       & --          & \textbf{0.44} & 0.72 & \underline{0.18} & 0.15 & 0.13 & 0.12
                          & 0.65 & \textbf{0.83} & 0.27 & 0.11 & 0.13 & 0.09 \\
    DisEnt    & --          & 0.42 & \textbf{0.73} & \textbf{0.18} & 0.15 & \underline{0.13} & \underline{0.12}
                          & \underline{0.65} & 0.83 & 0.26 & 0.12 & 0.13 & 0.09 \\
    \bottomrule
  \end{tabular}
  \label{tab:wavlm_debias_results}
  \vspace{-2mm}
\end{table*}

\subsection{\textbf{Regularization techniques}}
\label{subsec: regularization}
\subsubsection{\textbf{Gap regularization (GR)}}
\textbf{GR} is proposed in this paper to reduce disparities in model performance across demographic groups. We design an auxiliary loss to penalize discrepancies in multi-label fairness metrics in section~\ref{subsec:evaluation_metrics}, the  TPR\( _{gap}\) and FPR\(_{gap}\), as shown in Table~\ref{tab:debias_methods}. 
As training proceeds, the optimizer is encouraged to adjust feature representations and decision boundaries so that TPR and FPR become more similar between subpopulations. 
Because gap regularization directly targets group‐level disparities rather than merely re‐weighting individual samples, it is especially effective when some biases arise from unequal class‐conditional statistics (e.g., certain emotions co‐occur more frequently with one gender). 
We set the hyperparameter $\lambda_{GR}$ to 4.

\subsubsection{\textbf{Low Variance Regularization (LVR)}}
\textbf{LVR} \cite{LVR} enhances generalization by minimizing intra-class variance in the feature space. The method enforces embeddings of samples within the same emotion class to cluster around a dynamically computed class center via exponential moving average. This approach reduces intra-class variability and promotes robustness.

We extend LVR to multi-label SER by modifying the batch-wise class center computation:
\begin{equation}
    \bm{c_i^b} = (1-\omega) \frac{1}{B} \sum_{l=1}^{B} y_{l,i}\cdot\bm{h_{l,i}} + \omega \cdot \bm{c_i^{b-1}}, 
\end{equation}
where \(B\) is batch size, \( \omega\) is a hyperparameter regulating the influence of previous batch, \(y_{l,i}\) is the distributional label for \(l\)-th sample and class \(i\), and \(\bm{h_{l,i}}\) is the corresponding sample embedding. The multi-label regularization loss is then defined as:
\begin{equation}
    \mathcal{L}_r = \sum_{i=1}^k\sum_{l=1}^B y_{l,i}||\bm{h_{l,i}}-\bm{c_i^b}||^2.
\end{equation}

In addition, we integrate the computed class centers $c_i^b$ into the classification task by using them as auxiliary inputs, and we denote the corresponding auxiliary classification loss as \(\mathcal{L}_c\). This auxiliary loss provides structured class distribution information to improve performance. We use hyperparameter \(\lambda_{LVR} = 0.1\) and \(\omega=0.3\), following previous work \cite{LVR}.

\begin{table*}[t]
  \centering
  \sisetup{table-format=1.2}
  \caption{Overall performance of baseline and de‐biasing methods on MSP-PODCAST and CREMA-D using \textbf{XLSR upstream}. In each column, the best performance is shown in \textbf{bold} and the second‐best is \underline{underlined}.}
  \vspace{-2mm}
  \label{tab:debias_xlsr}
  \begin{tabular}{l c *{6}{S} *{6}{S}}
    \toprule
    \multirow{2}{*}{\textbf{Method}} 
      & \multirow{2}{*}{\textbf{BS}} 
      & \multicolumn{6}{c}{\textbf{MSP-PODCAST}} 
      & \multicolumn{6}{c}{\textbf{CREMA-D}} \\
    \cmidrule(lr){3-8} \cmidrule(lr){9-14}
      &  & {F1$\uparrow$} & {ACC$\uparrow$} & {TPR$_\mathrm{gap}$$\downarrow$} & {FPR$_\mathrm{gap}$$\downarrow$} & {F1$_\mathrm{gap}$$\downarrow$} & {DP$_\mathrm{gap}$$\downarrow$}
         & {F1$\uparrow$} & {ACC$\uparrow$} & {TPR$_\mathrm{gap}$$\downarrow$} & {FPR$_\mathrm{gap}$$\downarrow$} & {F1$_\mathrm{gap}$$\downarrow$} & {DP$_\mathrm{gap}$$\downarrow$} \\
    \midrule
    ERM       & --          & 0.41 & 0.72 & 0.18 & 0.14 & 0.13 & 0.12  & 0.68 & 0.83 & 0.25 & 0.12 & 0.12 & 0.09 \\ \midrule
    ADV       & \checkmark  & 0.37 & 0.74 & 0.53 & 0.31 & 0.48 & \textbf{0.04}
                          & \underline{0.67} & \underline{0.83} & 0.68 & 0.18 & 0.68 & 0.07 \\
    MADV      & \checkmark  & 0.37 & 0.70 & 0.49 & 0.28 & 0.45 & 0.10
                          & 0.66 & 0.83 & 0.67 & 0.18 & 0.67 & 0.09 \\
    GR        & \checkmark  & \underline{0.38} & 0.73 & 0.15 & 0.11 & 0.11 & 0.10
                          & 0.64 & 0.82 & 0.22 & 0.08 & 0.14 & 0.07 \\
    DS        & \checkmark  & 0.35 & \underline{0.74} & \underline{0.10} & \underline{0.04} & \underline{0.09} & 0.07
                          & 0.62 & 0.81 & \textbf{0.11} & \textbf{0.06} & \underline{0.07} & \textbf{0.04} \\
    RW        & \checkmark  & 0.36 & \textbf{0.75} & \textbf{0.07} & \textbf{0.04} & \textbf{0.09} & \underline{0.05}
                          & \textbf{0.67} & \textbf{0.83} & \underline{0.14} & \underline{0.07} & 0.08 & \underline{0.05} \\
    BLIND+d   & \checkmark  & 0.39 & 0.73 & 0.12 & 0.12 & 0.09 & 0.10
                          & 0.46 & 0.73 & 0.52 & 0.40 & 0.34 & 0.25 \\
    GDRO      & \checkmark  & 0.39 & 0.70 & 0.16 & 0.14 & 0.10 & 0.11
                          & 0.65 & 0.81 & 0.14 & 0.10 & 0.07 & 0.07 \\
    GADRO     & \checkmark  & \textbf{0.42} & 0.70 & 0.14 & 0.12 & 0.10 & 0.10
                          & 0.65 & 0.81 & 0.14 & 0.10 & \textbf{0.07} & 0.07 \\
    \midrule
    BLIND-d   & --          & \underline{0.40} & 0.71 & 0.17 & 0.16 & \underline{0.10} & 0.12
                          & 0.62 & 0.82 & \underline{0.20} & \textbf{0.09} & 0.11 & \textbf{0.07} \\
    LfF       & --          & 0.39 & \textbf{0.74} & \underline{0.14} & \textbf{0.10} & 0.12 & \underline{0.09}
                          & \textbf{0.67} & \underline{0.83} & 0.25 & 0.11 & 0.12 & 0.09 \\
    LVR       & --          & \textbf{0.42} & 0.66 & 0.25 & 0.19 & 0.14 & 0.15
                          & 0.66 & 0.81 & \textbf{0.17} & 0.10 & \textbf{0.08} & 0.09 \\
    SiH       & --          & 0.40 & 0.73 & 0.15 & 0.13 & 0.11 & 0.10
                          & \underline{0.67} & \textbf{0.83} & 0.27 & \underline{0.09} & 0.16 & 0.08 \\
    DisEnt    & --          & 0.36 & \underline{0.74} & \textbf{0.13} & \underline{0.11} & \textbf{0.10} & \textbf{0.09}
                          & 0.66 & 0.83 & 0.21 & 0.10 & \underline{0.09} & \underline{0.08} \\
    \bottomrule
  \end{tabular}
  \label{tab:xlsr_debias_results}
  \vspace{-2mm}
\end{table*}

\section{Experimental Results}

Since gender annotations may not always be available in real-world applications, we analyze two scenarios: one with demographic information (denoted as bias supervision, \textbf{BS}) and one without it. We adopt the Empirical Risk Minimization (\textbf{ERM}) model (\textbf{without applying any debiasing approaches}) as our primary baseline. All models are trained until convergence.

\definecolor{viridisA}{HTML}{E28743} 
\definecolor{viridisB}{HTML}{3da5c5} 
\definecolor{viridisC}{HTML}{b40f20} 
\definecolor{viridisD}{HTML}{063970} 

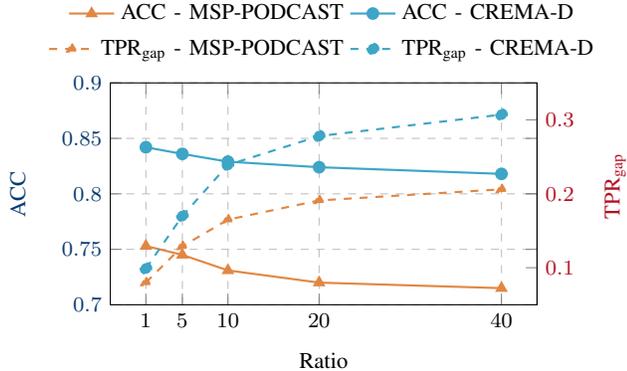
\begin{figure}[t]
\centering
\begin{tikzpicture}
  \tikzstyle{every node}=[font=\small]
  \begin{axis}[
      width=0.40\textwidth,
      height=0.25\textwidth,
      xlabel={Ratio},
      ylabel={ACC},
      y label style={color=viridisD},
      ymin=0.70, ymax=0.90,
      xtick={1,5,10,20,40},
      xmajorgrids=true,
      ymajorgrids=true,
      yticklabel style={color=viridisD},
      grid style={dashed},
      legend style={
        at={(0.5,1.22)},
        anchor=south,
        legend columns=2,
        draw=none,
        font=\small
      }
  ]
    \addplot[
      color=viridisA,
      mark=triangle*,
      thick
    ] coordinates {
      (1,0.753)
      (5,0.745)
      (10,0.731)
      (20,0.72)
      (40,0.715)
    };
    \addlegendentry{ACC - MSP-PODCAST}
    
    \addplot[
      color=viridisB,
      mark=*,
      thick
    ] coordinates {
      (1,0.842)
      (5,0.836)
      (10,0.829)
      (20,0.824)
      (40,0.818)
    };
    \addlegendentry{ACC - CREMA-D}
  \end{axis}
  
  \begin{axis}[
      width=0.40\textwidth,
      height=0.25\textwidth,
      xlabel={}, 
      ylabel={TPR$_\text{gap}$},
      yticklabel pos=right,
      ylabel style={
        at={(axis description cs:1.40,0.35)},
        anchor=west,
        color=viridisC
      },
      xtick={1,5,10,20,40},
      ymin=0.05, ymax=0.35,
      axis y line*=right,
      axis x line=none,
      yticklabel style={color=viridisC},
      legend style={
        at={(0.5,1.06)},
        anchor=south,
        legend columns=2,
        draw=none,
        font=\small
      }
  ]
    \addplot[
      color=viridisA,
      mark=triangle*,
      thick,
      dashed
    ] coordinates {
      (1,0.080)
      (5,0.129)
      (10,0.165)
      (20,0.191)
      (40,0.206)
    };
    \addlegendentry{TPR$_\text{gap}$ - MSP-PODCAST}
    
    \addplot[
      color=viridisB,
      mark=*,
      thick,
      dashed
    ] coordinates {
      (1,0.099)
      (5,0.170)
      (10,0.239)
      (20,0.278)
      (40,0.307)
    };
    \addlegendentry{TPR$_\text{gap}$ - CREMA-D}
  \end{axis}
\end{tikzpicture}
\vspace{-2mm}
\caption{Hamming accuracy (\textbf{ACC}) and TPR$_{gap}$ under various gender-biased data distribution conditions. The X-axis represents the \textbf{Ratio}.}
\label{fig:ratio_plot}
\vspace{-1mm}
\end{figure}

\subsection{Impacts of Data Distribution}
The effects of the data distribution on classification performance and bias in SER systems are presented in Table~\ref{tab:detailed_ratios_results} and Fig.~\ref{fig:ratio_plot}. 
Accuracy declines consistently with increasing gender imbalance, reaching its lowest in a ratio of 1:40, indicating reduced generalization between genders. Simultaneously, TPR$_{gap}$ rises steadily, reflecting a growing disparity in TPR between male and female speakers. This suggests that as the training set becomes more skewed, the model increasingly favors the majority gender, amplifying performance bias. 
\textbf{The following sections evaluate debiasing methods under a 1:20 imbalance condition}, where both accuracy degradation and fairness disparities are significant.

\subsection{Limitation of Single-Label Adversarial Debias in Multi-Label SER}
Previous debiasing literature on single-label SER tasks \cite{Chien_2024, Upadhyay_2025} consistently shows the effectiveness of adversarial approaches, such as \textbf{ADV}. 
However, surprisingly, the popular adversarial training strategy showed mixed results in our multi-label SER context. 
ADV significantly reduces the \textbf{DP$_{gap}$}, but at the cost of worsening other fairness gaps and harming overall SER accuracy.
This suggests that adversarial debiasing, while effective in single emotion SER, struggles in multi-label settings where emotions frequently co-occur. In multi-label SER, each input can trigger multiple correlated emotions, so forcing the adversary to remove gender information can unintentionally disrupt the subtle relationships among emotions.

\subsection{Robustness and Effectiveness of Debiasing Methods}



In Table~\ref{tab:wavlm_debias_results}, eight out of the thirteen experimental results (specifically, \textbf{GR, DS, RW, GADRO, LfF, LVR, SiH,} and \textbf{DisEnt}) successfully reduced bias across all fairness metrics in both databases, compared to the baseline model, \textbf{ERM}. 
Similarly, in Table~\ref{tab:xlsr_debias_results}, six out of the thirteen experimental results (\textbf{DS, RW, GDRO, GADRO, LfF,} and \textbf{DisEnt}) also demonstrated a reduction in bias across all fairness metrics in both databases. 
Among these, \textbf{GADRO} (with bias supervision) and \textbf{LfF} (without bias supervision) consistently outperform ERM on all four fairness metrics for both WavLM and XLSR, while incurring only minimal drops in F1 and ACC. 
This dual requirement—better fairness across two datasets and two backbones, plus negligible performance loss—\textbf{makes GADRO and LfF the most robust debiasing methods under high gender imbalance.} 


\subsection{Findings of Debiasing Methods with BS} 
In most real applications, we prefer to maintain similar accuracy in SER while reducing performance bias between female and male speakers, compared to the \textbf{ERM}. Based on this criterion, \textbf{RW} emerges as the most effective debiasing method. It consistently obtains a macro-F1 score comparable to that of the baseline ERM while reducing the bias. In contrast, while \textbf{DS}  often records the lowest gap values (especially in $\text{TPR}_{gap}$ and $\text{FPR}_{gap}$), it degrades SER performance. This highlights the inherent trade-off between optimizing for pure recognition accuracy and enforcing fairness constraints. 

\subsection{Findings of Debiasing Methods without BS}
Among the methods without BS, \textbf{LVR} and \textbf{SiH} achieve the most favorable trade‐off between recognition accuracy and fairness when compared to the \textbf{ERM} baseline. In particular, \textbf{LVR} consistently produces some of the lowest $\text{TPR}_{gap}$ and $\text{FPR}_{gap}$ while maintaining overall ACC and F1 very close to those of ERM. Across WavLM and XLSR experiments, LVR yields the largest bias reduction in seven of the sixteen measured bias metrics, demonstrating its ability to minimize performance disparities without sacrificing recognition performance. \textbf{SiH} tends to deliver the highest F1 in several cases, although its fairness metrics are not always as low as those of LVR. In contrast, \textbf{BLIND‐d} shows performance curves almost identical to ERM, indicating only marginal improvements in fairness.

Also, debiasing methods without BS consistently leave larger performance disparities than methods that use gender labels. For example, in Table~\ref{tab:wavlm_debias_results} (WavLM upstream), the best $\text{TPR}_{gap}$ and $\text{FPR}_{gap}$
among methods with BS are 0.08 and 0.06 (achieved by RW), whereas the best method without BS still has $\text{TPR}_{gap}$ and $\text{FPR}_{gap}$
of 0.18 and 0.13 (achieved by LVR), respectively. These results show that, without explicit gender labels, fairness objectives are inherently harder to optimize, resulting in higher bias gaps across all non–BS methods.

\section{Conclusion}
We introduced a comprehensive benchmark, \textbf{EMO-Debias}, for debiasing multi-label SER systems in this work. Our evaluation spans the proposed GR and 12 other adapted debiasing methods. 
Eight of them employ explicit bias supervision, and five operate without it. 
Our assessment reveals three key findings: 
(1) The data distribution of gender plays a crucial role in shaping both SER performance and bias; imbalanced training data leads to marked disparities, underscoring the importance of balanced datasets or robust debiasing strategies. 
(2) Among methods that use gender labels, \textbf{GADRO} and \textbf{RW} stand out for reducing bias across all fairness metrics on both backbones and datasets while incurring only minimal macro-F1 scores and accuracy drops; by contrast, adversarial approaches often trade one type of fairness gap for others.
(3) For scenarios without bias supervision (i.e., demographic information such as gender), our modified \textbf{LVR} technique achieves the best trade‐off for mitigating gender bias while maintaining strong SER performance.

\section{Limitation and Future Work}
The current work assumes gender as a binary attribute (male/female), which does not capture the full spectrum of gender identities \cite{10832234}. While our study primarily focuses on gender bias, we recognize that other factors, such as age, cultural background, and language, could also influence SER disparities \cite{10832317, 10832259}. Besides, due to limited variability in public emotion databases, we simulated imbalanced conditions that might not fully reflect real-world data. Future work will incorporate additional bias dimensions by evaluating our methods on more diverse, multilingual databases to better understand demographic influences and develop more equitable and fair SER systems.


\bibliographystyle{IEEEtran}
\bibliography{IEEEref}

\end{document}